\documentclass[12pt]{article}
\usepackage{amssymb}
\usepackage{amsmath}
\usepackage{epsfig}
\usepackage{float}
\newcommand{\be}{\begin{equation}}
\newcommand{\ee}{\end{equation}}
\newcommand{\bea}{\begin{eqnarray}}
\newcommand{\eea}{\end{eqnarray}}
\begin{document}

\begin{center}
{\bf FLAVOR NEUTRINO OSCILLATIONS AND TIME-ENERGY UNCERTAINTY RELATION}\footnote{ A report at the 2nd Scandinavian Neutrino Workshop, SNOW 2006, Stockholm, May 2-6, 2006.}

\end{center}

\begin{center}
S. M. Bilenky
\end{center}

\begin{center}
{\em  Joint Institute
for Nuclear Research, Dubna, R-141980, Russia, and\\
SISSA,via Beirut 2-4, I-34014 Trieste, Italy.}
\end{center}

\begin{abstract}
We consider neutrino oscillations as non stationary phenomenon based on Schrodinger evolution equation 
and mixed states of neutrinos with definite flavors.
We show that time-energy uncertainty relation plays a crucial role
in neutrino oscillations.  We compare neutrino oscillations with
$B_{d}^{0}\leftrightarrows\bar B_{d}^{0}$  oscillations.
\end{abstract}
\section{Introduction}
The recent accelerator K2K \cite{K2K} and MINOS \cite{Minos} experiments
is an important step in the study of the problem of neutrino masses
and mixing. In these experiments with neutrino beams fully under
control neutrino oscillations, discovered in the
Super-Kamiokande \cite{SK}, SNO \cite{SNO}, KamLAND \cite{Kamland}
and other neutrino experiments \cite{Gallex,Sage,SKsol}, were
confirmed.

I would like to stress that K2K  and MINOS experiments are {\em neutrino
oscillation experiments of a new type}. In these experiments for the
first time it was measured time of neutrino production and time of neutrino detection.

In the K2K experiment neutrinos were produced in 1.1 $\mu\rm{sec}$
spills. 
Neutrino events which satisfy the criteria
\begin{equation}\label{1}
-0.2 \leq [(t_{SK} -t_{KEK})- t_{TOF}]\leq 1.3~ \mu \rm{sec}
\end{equation}
were selected. Here $t_{KEK}$ is the time of the production of neutrinos at KEK, $t_{SK}$ is the time of 
the detection of neutrinos in the Super-Kamiokande detector and 
 $ t_{TOF}\simeq L/c $ is the time of
flight. In the K2K experiment $L\simeq 250$ km and $ t_{TOF} \simeq
0.83\cdot 10^{3}\mu\rm{sec}$.

In spite neutrino oscillations are discovered, there  exist
different interpretation of this new phenomenon. (see, for example,
\cite{Carlo} and references therein).  From our point of view K2K and
MINOS experiments are important for the understanding of the
physics of neutrino oscillations \cite{Bil,BilMat}.
In these accelerator experiments it was shown
 that neutrino
oscillations is phenomenon with {\em finite time interval} $\Delta
t$ during which neutrino state is significantly changed (for
example, $\nu_{\mu}$ is transferred into superposition of
$\nu_{\mu}$ and $\nu_{\tau}$). For such phenomena
time-energy uncertainty relation takes place (see, for example, \cite{Messiah,Sakurai,Bauer})
\begin{equation}\label{2}
\Delta E~ \Delta t \geq 1,
\end{equation}
where $\Delta E$ is uncertainty in energy. This means that 
neutrinos are described by {\em superpositions of states with different energies} (nonstationary states).

Neutrino oscillations and
$K^{0}\leftrightarrows \bar K^{0}$, $B^{0}_{d}\leftrightarrows \bar
B^{0}_{d}$ oscillations  have the same
quantum-mechanical origin. This was original idea of B. Pontecorvo who was pioneer of neutrino
oscillations hypothesis \cite{Pont}. Recently in high-precision  experiments at asymmetric B-factories
$B^{0}_{d}\leftrightarrows \bar
B^{0}_{d}$ oscillations  were studied in details. Nonstationary picture of oscillations
was perfectly confirmed.
We will discuss here the formalism which is common for  flavor oscillations of neutral mesons and flavor neutrino oscillations.
\section{Field theoretical basics}
From the point of view of the field theory neutrino oscillations are based on the following assumptions.
\begin{itemize}
\item
Lagrangians of interaction of neutrinos are given by the SM
\begin{equation}
\mathcal{L}_{I}^{\mathrm{CC}} = - \frac{g}{2\sqrt{2}} \,
j^{\mathrm{CC}}_{\alpha} \, W^{\alpha} + \mathrm{h.c.};~~
\mathcal{L}_{I}^{\mathrm{NC}} = - \frac{g}{2\cos\theta_{W}} \,
j^{\mathrm{NC}}_{\alpha} \, Z^{\alpha}.\label{3}
\end{equation}
Here $g$ is the $SU(2)$ gauge coupling, $\theta_{W}$ is the weak
angle and
\begin{equation}
j^{\mathrm{CC}}_{\alpha} =2 \sum_{l=e,\mu,\tau}  \bar \nu_{lL}
\gamma_{\alpha}l_{L};~~ j ^{\mathrm{NC}}_{\alpha}
=\sum_{l=e,\mu,\tau} \bar \nu_{lL}\gamma_{\alpha}\nu_{lL}\label{3a}
\end{equation}
are charged and neutral currents.
\item
Neutrino masses are different from zero and 
fields of neutrinos with definite masses enter into CC and NC in a mixed form
\begin{equation}\label{4}
\nu_{l}=\sum_{i}U_{li}~\nu_{i}.
\end{equation}
Here $U^{\dagger}U=1$ and $\nu_{i}$ is a field of neutrino with mass $m_{i}$
\item
A  neutrino mass term enters into the total Lagrangian. The type of neutrino mass term  at present is 
unknown. There are two completely different possibilities: Majorana mass term (in case if there are no conserved
lepton numbers)
or Dirac mass term (if total lepton number is conserved).
\end{itemize}
\section{Flavor neutrino states}
Flavor neutrinos $\nu_{e},\nu_{\mu},\nu_{\tau}$ which are produced and detected in CC
weak processes together with corresponding leptons, are described by {\em  mixed states}.
Let us consider, for example, the decay
\begin{equation}\label{5}
a \to b + l^{+}+ \nu_{i}
\end{equation}
From analysis of neutrino oscillation data \cite{SK, SNO, Kamland,
Gallex,Sage,SKsol} for the neutrino mass-squared differences the following best-fit values were found:
$\Delta m^{2}_{23} \simeq 2.4
\cdot 10^{-3} \rm{eV}^{2}$ and $\Delta m^{2}_{12} \simeq 8.0 \cdot
10^{-5} \rm{eV}^{2}$. 
These values are so small that due to Heisenberg
uncertainty relation it is impossible to distinguish 
momenta of produced
neutrinos with different masses (see, for example, \cite{Bil}).
For the state of the final
neutrinos we have
\begin{equation}\label{6}
|\nu_{f}\rangle =  \sum_{i}|\nu_i\rangle  ~ \langle \nu_i
\,l^{+}\,b\,| S |\,a\rangle,
\end{equation}
where $|\nu_i\rangle $ is the state of neutrino with mass $m_{i}$
and  momentum $\vec{p}$ and $ \langle \nu_i, \,l^{+}\,b\,|
S |\,a\rangle $ is the matrix element of the process (\ref{5}).

Neutrino energies in neutrino experiments ($\gtrsim $ MeV) are much
larger than neutrino masses ($\lesssim $ 1 eV).
 Thus, neutrino masses can be neglected in the matrix element. 
We have
\begin{equation}\label{7}
\langle \nu_{i}\,l^{+}\,b\,| S |\,a\rangle \simeq U_{l i}^{*}\,~
\langle \nu_{l}\,l^{+}\,b\,| S |\,a\rangle_{SM},
\end{equation}
where $\langle \nu_{l}\,l^{+}\,b\,| S |\,a\rangle_{SM}$ is the SM matrix element of
the process
$a \to b + l^{+}+ \nu_{l}$.
For the final neutrino state $|\nu_{f}\rangle$ we have
\begin{equation}\label{8}
|\nu_{f}\rangle =  |\nu_{l}\rangle ~\langle \nu_{l}\,l^{+}\,b\,| S
|\,a\rangle_{SM}.
\end{equation}
Here
\begin{equation}\label{9}
|\nu_{l}\rangle=\sum_{i}U^{*}_{li}~|\nu_{i}\rangle
\end{equation}
is the  normalized state of flavor neutrino $\nu_{l}$ (produced in CC weak process
together with $l^{+}$)

Thus, because of the  smallness of neutrino mass-squared differences
\begin{itemize}
\item
In production (and detection) processes flavor lepton numbers are effectively conserved.
\item
Matrix elements of neutrino production and detection processes are
given by the SM.
\end{itemize}
In strong interaction quark flavor is conserved. Therefore, in decays of
$\Upsilon (4S)$ flavor $B_{d}$ and $\bar B_{d}$ mesons  are
produced. These particles are described by mixed states
\begin{equation}\label{10}
|B_{d}\rangle=\frac{1}{2N}~(|B_{H}\rangle+|B_{L}\rangle;~
|\bar
B_{d}\rangle=\frac{1}{2N}~\frac{p}{q}~(|B_{H}\rangle-|B_{L}\rangle),
\end{equation}
where $|B_{H,L}\rangle$ are states of neutral B-mesons with definite
masses and widths, $N$ is the normalization factor,  
$p=\sqrt{H_{B_{d}\bar B_{d}}}$ and $q=\sqrt{H_{\bar B_{d} B_{d}}}$.
The relations
(\ref{9}) are neutrino analogy of the relations   (\ref{10}).

\section{Mixed neutrino states and invariance under translation}
Let us consider translations in space and time
\begin{equation}\label{11}
x'_{\alpha} = x_{\alpha} +a_{\alpha},
\end{equation}
where $a_{\alpha}$ is arbitrary vector. In the case of the
invariance under tarnslations we have
\begin{equation}\label{12}
|\Psi \rangle' = e^{i\,P\, a }\, |\Psi \rangle.
\end{equation}
Here $P_{\alpha}$ is the operator of the total momentum and vectors
$|\Psi \rangle $ and $|\Psi \rangle'$ describe the {\em the same
state.} If,  for example, $|\Psi \rangle$ is the state with total
momentum $p$ the state $|\Psi \rangle'$ differs from   $|\Psi
\rangle$ by the phase factor
\begin{equation}\label{13}
|\Psi \rangle' = e^{i\,p\, a }\, |\Psi \rangle,
\end{equation}

In the case of the mixed flavor neutrino states we have
\begin{equation}\label{13a}
|\nu_{l}\rangle'= e^{i\,P\, a }\, |\nu_{l}\rangle =
e^{-i\,\vec{p}\,\vec{a}}\,\sum_{l'} |\nu_{l'}\rangle \,\sum_{i}U_{l'
i}e^{i\,E_{i}\,a^{0}}\,U_{l i}^*
\end{equation}
Thus,  vectors $|\nu_{l}\rangle'$ and    $|\nu_{l}\rangle$   describe {\em different
states}. We come to the conclusion that in the case of the mixed flavor neutrino  states
there is no invariance under translation in time. This corresponds to time-energy uncertainty relation and
nonconservation of energy in
neutrino oscillations.

\section{Evolution of mixed flavor neutrino states}
Evolution equation in the quantum field theory is Schrodinger equation
\begin{equation}\label{14}
i\,\frac{\partial |\Psi(t) \rangle}{\partial t} = H\, |\Psi(t) \rangle.
\end{equation}
If at $t=0$ flavor neutrino $\nu_{l}$ described by the mixed flavor
state $|\nu_{l}\rangle$  is produced, from
(\ref{14}) it follows that at the time $t$ the neutrino  is
described by {\em non stationary state}
\begin{equation}\label{15}
|\nu_{l}\rangle_{t} =\sum_{i}e^{-iE_{i}t}U_{l
i}^*|\nu_{i}\rangle=\sum_{l'}|\nu_{l'}\rangle~ \sum_{i}U_{l'i}~
e^{-iE_{i}\,t}\,U_{l i}^*.
\end{equation}
In the case of  $B_{d}$ and  $\bar B_{d}$ mesons
analogous non stationary states are given by
\begin{equation}\label{15a}
|B_{d}\rangle_{t}=g_{+}(t)~|B_{d}\rangle +\frac{q}{p}~g_{-}(t)~|\bar
B_{d}\rangle;~~ |\bar
B_{d}\rangle_{t}=g_{-}(t)~\frac{p}{q}~|B_{d}\rangle +g_{+}(t)~|\bar
B_{d}\rangle.
\end{equation}
Here
\begin{equation}\label{16}
g_{\pm}(t)=\frac{1}{2}~(e^{-i\lambda_{H}t}\pm e^{-i\lambda_{L}t});~~
\lambda_{H,L}=m_{H,L}-\frac{i}{2}~\Gamma_{H,L},
\end{equation}
where $m_{H,L}$ and $\Gamma_{H,L}$ are masses and total decay widths
of $B_{H,L}$-mesons.  All successful $B_{d}- \bar B_{d}$ phenomenology is based on these
relations.

From  (\ref{15}) it follows that the state of neutrino is
significantly changed at the time $t$, which satisfies (at least for 
one $\Delta m^{2}_{ik}$ ($i\not=k$)) the  inequality
\begin{equation}\label{17}
(\Delta E)_{ik}~t =\frac{\Delta m^{2}_{ik}}{2E}~t \geq 1.
\end{equation}

If in accordance with the results of K2K and MINOS experiments we put
$t\simeq L$ from (\ref{16}),  we obtain  the standard expression for
$\nu_{l}\to\nu_{l'} $ transition probability
\begin{equation}\label{18}
{\mathrm P}(\nu_{l} \to \nu_{l'}) =|\delta_{l'l}+
\sum_{i>1}U_{l' i} \,~( e^{- i\,\Delta m^2_{1i} \frac {L}{2E}}-1) \,~U^*_{l i} \, |^2
\end{equation}
It is obvious that the time-energy uncertainty relation  (\ref{17})
is well known necessary condition to observe neutrino oscillations
\cite{BilPont}:
\begin{equation}\label{19}
\frac{\Delta m^{2}_{ik}}{2E}~L \geq 1.
\end{equation}
We assumed that flavor neutrino state is
determined by momentum and energies of states with different masses are different. 
Let us notice that in approach based on
Schrodinger evolution equation this is the only possibility
compatible with experimental data (otherwise in oscillation phases 
arbitrary additional terms appear).

It was  stated in literature
 (see \cite{Kayser, Stodol,Lipkin}) that   in neutrino oscillations time is not measured and
only distance $L$ is relevant. In this approach oscillation
probabilities are averaged over time and as a result energies of
different mass components are the same (and momenta are different).
In spite final expression for the transition probability obtained in
\cite{Kayser, Stodol,Lipkin} coincides with the standard one (Eq.
(\ref{18}) this approach to neutrino oscillations does not
corresponds to the results of the accelerator long baseline experiments.

We will present another argument against ``equal energy assumption''
. Let us consider the standard effective Hamiltonian of neutrino in
matter \cite{Wolf}
\begin{equation}\label{20}
\mathrm{H}= \mathrm{H_{0}}+\mathrm{H_{I}}
\end{equation}
Here
\begin{equation}\label{21}
(\mathrm{H_{I}})_{l'l}=\sqrt{2}\,G_{F}\,\rho_{e}\,\eta_{l'l}
\end{equation}
is effective Hamiltonian of neutrino in matter ($\rho_{e}$ is electron number density and $\eta_{ee}=1$)
and $\mathrm{H_{0}}$ is the free Hamiltonian.
In the flavor representation  we have
\begin{equation}\label{22}
(\mathrm{H_{0}})_{l'l}   =\langle \nu_{l'}|H_{0}|\nu_{l}\rangle = \sum_{i}
U_{l'i}~E_{i}~U^{*}_{li}
\end{equation}
If we assume that $E_{i}=E$ in this case free Hamiltonian is an unit
matrix $ (\mathrm{H_{0}})_{l'l} =E\delta_{l'l}$. It will be  no
matter MSW effect in this case in contradiction to the results of
the solar neutrino experiments  (see \cite{Lisi}).
\section{Conclusion}
The understanding of the physics of neutrino oscillations is an important issue.
 Accelerator K2K and MINOS experiments
are new type of neutrino oscillation experiments: in these
experiments time of neutrino production and neutrino detection  was
measured. The finite time during which neutrino state is
significantly changed according to time-energy uncertainty relation
requires uncertainty of energy. This means that neutrino is
described by non stationary state.

Neutrino oscillations and $B_{d}^{0}\leftrightarrows\bar B_{d}^{0}$
etc oscillations have the same origin.  We present here the
formalism which in its basics is common for neutrino oscillations
and flavor oscillations of neutral mesons.

I  would like to acknowledge the Italian program ``Rientro dei
cervelli'' for the support.

\end{document}